\documentclass[a4paper,english]{article}
\usepackage[T1]{fontenc}
\usepackage[latin9]{inputenc}
\makeatletter

\newcommand{\lyxaddress}[1]{
\par {\raggedright #1
\vspace{1.4em}
\noindent\par}
}

\usepackage{babel}
\makeatother

\begin{document}

\title{Anomalous delta-type electric and magnetic two-nucleon interactions}

\author{Nicolae Bogdan Mandache%
\thanks{email: mandache@infim.ro%
}}

\date{{}}

\maketitle

\lyxaddress{National Institute for Lasers, Plasma and Radiation Physics, Magurele-Bucharest,
POBox MG-36, Romania}

\begin{abstract}
Anomalous delta-type interactions, of both electric and magnetic nature,
are introduced between the overlapping peripheral structures of the
nucleons, which may explain the spin-triplet deuteron state and the
absence of other nucleon-nucleon bound states. 
\end{abstract}


As evidenced by deuteron, the $p-n$ interaction favours the coupling
of the proton and neutron intrinsic spins to \emph{S=1} (lower energy)
rather than $S=0$, which is in striking contrast to the like-nucleon
residual interaction (pairing) that favours $S=0$.\cite{key-1} As
it is well known, a spin dependent nuclear potential is introduced
(the triplet potential is greater in absolute value than the singlet
potential) to explain why the $n-p$ triplet state is bound in contrast
to unbound $n-p$, $n-n$ and $n-p$ singlet states, which eventually
solves the above contradiction.\cite{key-1}

The magnetic momentum of the proton ($+2.79\mu_{N}$, $\mu_{N}$ being
the nuclear magneton) and the magnetic momentum of the neutron ($-1.91\mu_{N}$)
exhibit in deuteron an \char`\"{}antiferromagnetic\char`\"{} alignment.
This is in contrast with the hydrogen atom, where, by hyperfine interaction,
the magnetic momentum of the electron ($-1\mu_{B}$, $\mu_{B}$ being
the Bohr magneton) gets parallel with the magnetic momentum of the
proton, and the ground-state of the hydrogen is a spin singlet (para-hydrogen).

We suggests in this short letter a possible explanation for such a
discrepancy, which may also throw light on the inexistence of other
nucleon-nucleon bound states than the deuteron. 

According to the laws of the electromagnetism and to our classical
representation of the magnetic (and electric) momenta, it is very
likely that the magnetic interaction is ferromagnetic at large distances,\emph{
i.e.} it favours the parallel alignment of the magnetic momenta. Similarly,
by the same arguments, it is very likely that this interaction is
antiferromagnetic, \emph{i.e.} it favours the anti-parallel alignment
of the magnetic momenta, at very short distances. It looks therefore
as a delta-type interaction. 

For the electron the ferromagnetic alignment is favourable, even at
close range to the proton, in view of its point-like structure. This
may explain the spin-singlet ground-state of the hydrogen. 

For nucleons in nuclei, the situation is different. The spatial extension
of the nucleons is well documented, as it is their compact packing
in nuclei. A typical example for antiferromagnetic alignement could
be the pairing interaction. Every pair of like nucleons in the same
orbit (the wave functions of the two nucleons strongly overlap) couple
to spin zero and zero magnetic moment for the lowest energy state. 

There are some other puzzles related to this tight packing of the
nucleons in atomic nuclei, like, for instance, the difference in the
lifetimes of the free neutron and the bound neutron. 

In deuteron there is a substantial overlapping between the peripheral
nucleonic clouds put togheter around the central nucleonic cores,
reminiscent of the molecular orbitals,\cite{key-2} which favours
a strong residual delta-interaction. The high density cores of proton
and neutron placed each other at a distance of $2fm$, in the central
part of deuteron, interact electro-magnetically much weaker.

According to recent trends in conceiving the internal spin structure
of the nucleons, it seems very likely that the anomalous magnetic
momenta ($+1.79\mu_{N}$ for proton and $-1.91\mu_{N}$ for neutron)
are to a great extent peripherally distributed. The pion contribution
to the magnetic momenta of the nucleons was estimated to $+1\mu_{N}$
for proton and $-1\mu_{N}$ for neutron as arising in the peripheral
cloud.\cite{key-3,key-4} These anomalous magnetic momenta are therefore
very prone to undergo an antiferromagnetic delta-interaction, which
would explain thus the spin-triplet ground-state of the deuteron and
the absence of the spin-singlet proton-neutron bound state. 

At the same time, the electromagnetic nature of the magnetic momenta
suggests also a peripheral delta -type electric (coulomb) interaction
between the nucleons. The spatial extension of the nucleons imply
both a spatial distribution of electric charge and magnetic momenta.\cite{key-3}-\cite{key-7}
It is estimated that the negative charge in neutron extends from $1fm$
to $2-3fm$, the core being positively charged (such that the overall
charge vanishes).\cite{key-2,key-3} A similar but positive charge
extension holds for proton (it is worth recalling here the approximate
diameter of cca $4fm$ of the deuteron). 

Hence the delta-type magnetic interaction is attractive for the $n-p$
triplet and the $n-n$ and $p-p$ singlet states, and is repulsive
for the $n-p$ singlet state. The delta-type electric (coulomb) interaction
is attractive for the $n-p$ triplet and singlet states and repulsive
for the $n-n$ and $p-p$ singlet states. Consequently the most attractive
is the $n-p$ triplet state, the only one with both delta-type interactions
attractive.

Based on these observations and assuming that the nuclear interaction
is the same for all the four states we may express the nucleon-nucleon
interaction by means of three energies. The energy $E_{N}$ associated
with the strong nuclear force (nucleons kinetic energy minus the nuclear
potential well) and two anomalous delta-type interaction energies,
one electrical denoted by $E_{e}$ and another magnetic denoted by
$E_{m}$. We assume that $E_{m}$ and $E_{e}$, respectively, are
the same for the $n-p$, $n-n$ and $p-p$, taking into account that
the anomalous magnetic moments of proton and neutron are comparable
in absolute values. This means that the fractions of electric charges
and magnetic moments carried by the peripheral parts of proton and
neutron are approximately equal in absolute values and of opposite
signs.

The nucleon-nucleon relevant energies are the binding energy $-2.22MeV$
for the triplet proton-neutron state (deuteron), the energy $+0.07MeV$
for the singlet proton-neutron state and the energy $+0.15MeV$ for
the singlet neutron-neutron and proton-proton states (after removing
of course the proton-proton Coulomb repulsion between point-like charges).\cite{key-4,key-7}
We may write therefore:\[
\begin{array}{c}
-E_{e}-E_{m}+E_{N}=-2.22MeV\,\,(n-p\,\, triplet)\\
\\-E_{e}+E_{m}+E_{N}=0.07MeV\,\,(n-p\, simglet)\\
\\+E_{e}-E_{m}+E_{N}=0.15MeV\,\,(n-n\, and\, p-p\, singlets)\end{array}\]

From these equations we get $E_{e}=1.19MeV$, $E_{m}=1.15MeV$ and
$E_{N}=0.11MeV$. Since $E_{N}$ is positive it follows that all four
basic states are sligthly unbound by nuclear force (the kinetic energy
of the two nucleons slightly exceeds the nuclear potential well).
The two anomalous delta-type magnetic and electric interactions make
the difference between bound and unbound two-nucleon states.

The proton-proton and neutron-neutron magnetic delta-interaction introduced
here is analogous to the corresponding pairing interaction which couples
the like nucleons to the state of zero total angular momentum and
zero magnetic moment, like in magic nuclei for instance.\cite{key-1}
It is worth noting that the strength derived here for this magnetic
interaction, $E_{m}=1.15MeV$, is comparable in magnitude to the pairing
interaction (cca $1MeV$).

In conclusion, we put forward herein the hypothesis that anomalous
delta-type electric and magnetic interactions may arise between the
overlapping peripheral parts of the nucleons, and analyzed its conseqeunces.
It is shown that such a hypothesis may explain the deuteron bound
state and the inexistence of other nucleon-nucleon bound states, assuming
an equal nuclear interaction for all four states.

\textbf{Acknowledgments.} The author expresses his sincere thanks
to M. Apostol for valuable conversations and suggestions, and for
help in making a clear exposition of these ideas.

\end{document}